\documentclass[superscriptaddress,twocolumn,showpacs,preprintnumbers,amsmath,amssymb,prl]{revtex4}

\usepackage{graphicx}
\usepackage{dcolumn}
\usepackage{bm}
\usepackage{multirow}
\usepackage{color}

\begin{document}


\title{Spin Fluctuations from Hertz to Terahertz on a Triangular Lattice}

\author{Yusuke Nambu\footnote{Present address: Institute for Materials Research, Tohoku University, Sendai, Miyagi 980-8577, Japan}}
\affiliation{Institute of Multidisciplinary Research for Advanced Materials, Tohoku University, Sendai, Miyagi 980-8577, Japan}
\affiliation{Neutron Science Laboratory, Institute for Solid State Physics, University of Tokyo, Tokai, Ibaraki 319-1106, Japan}
\author{Jason S. Gardner}
\affiliation{NIST Center for Neutron Research, National Institute of Standards and Technology, Gaithersburg, Maryland 20899, USA}
\affiliation{National Synchrotron Radiation Research Center, Neutron Group, Hsinchu 30077, Taiwan}
\author{Douglas E. MacLaughlin}
\affiliation{Department of Physics and Astronomy, University of California, Riverside, California 92521, USA}
\author{Chris Stock}
\affiliation{School of Physics and Astronomy and Centre for Science at Extreme Conditions, University of Edinburgh, Edinburgh EH9 3FD, United Kingdom}
\author{Hitoshi Endo}
\affiliation{Quantum Beam Science Directorate, Japan Atomic Energy Agency, Tokai, Ibaraki 319-1195, Japan}
\author{Seth Jonas}
\affiliation{Institute for Quantum Matter and Department of Physics and Astronomy, Johns Hopkins University, Baltimore, Maryland 21218, USA}
\author{Taku J Sato}
\affiliation{Institute of Multidisciplinary Research for Advanced Materials, Tohoku University, Sendai, Miyagi 980-8577, Japan}
\author{Satoru Nakatsuji}
\affiliation{Institute for Solid State Physics, University of Tokyo, Kashiwa, Chiba 277-8581, Japan}
\affiliation{PRESTO, Japan Science and Technology Agency, Saitama 332-0012, Japan}
\author{Collin Broholm}
\affiliation{Institute for Quantum Matter and Department of Physics and Astronomy, Johns Hopkins University, Baltimore, Maryland 21218, USA}
\affiliation{NIST Center for Neutron Research, National Institute of Standards and Technology, Gaithersburg, Maryland 20899, USA}

\date{\today}

\begin{abstract}
The temporal magnetic correlations of the triangular lattice antiferromagnet NiGa$_2$S$_4$ are examined through thirteen decades ($10^{-13}-1$~sec) using ultra-high-resolution inelastic neutron scattering, muon spin relaxation, AC and nonlinear susceptibility measurements.
Unlike the short-ranged {\it spatial} correlations, the temperature dependence of the {\it temporal} correlations show distinct anomalies. 
The spin fluctuation rate decreases precipitously upon cooling towards $T^{\ast}=8.5$~K, but fluctuations on the microsecond time scale then persist in an anomalous dynamical regime for 4 K $<T\le T^{\ast}$.
As this time scale exceeds that of single site dynamics by six orders of magnitude, these fluctuations bear evidence of emergent degrees of freedom within the short-range correlated incommensurate state of NiGa$_2$S$_4$.
\end{abstract}


\maketitle

The two dimensional triangular lattice plays a prominent role in the field of frustrated magnetism.
While it now seems clear that the conventional triangular lattice antiferromagnet has long-range non-collinear 120 degree N\'{e}el order irrespective of the spin quantum number, $S$~\cite{triLRO1,triLRO2,triLRO3}, experiments on materials with generalized longer range interactions, spin-orbit interactions, as well as lattice distortions point to a rich variety of phases in the vicinity of the N\'{e}el phase \cite{organic1,organic2}.
In the quantum limit ($S=1/2$) first considered by Anderson \cite{RVB}, experiments on organic materials suggest a gapless spin liquid with a spinon Fermi surface may exist near the metal insulator transition \cite{QSL}.
Recent exact-diagonalization showed such quantum spin-liquid states can be stabilized by the presence of randomness in the near-neighbor interactions~\cite{RandomSinglet}.
Deep in the insulating limit, neutron scattering experiments suggest conventional spin waves may be supplemented or replaced by a spinon-like continuum even when long-range order is present \cite{BaCoSbO}.

For larger spin quantum numbers theoretical \cite{Z2} and experimental \cite{NaCrO2} work points to complex slow spin dynamics that is still poorly understood.
Here we focus on such dynamics in the $S=1$ triangular lattice antiferromagnet NiGa$_2$S$_4$~\cite{nigasScience,nigasJPSJ}, which is distinguished by a 2D incommensurate critical wave vector and the potential for spin-nematic interactions \cite{nematic1,nematic2}.
Using high energy resolution inelastic neutron scattering, neutron spin echo, muon spin relaxation ($\mu$SR), and AC magnetometry, we provide evidence for emergent MHz spin dynamics over a wide range of temperatures.
We argue such slow spin dynamics is associated with emergent topologically protected degrees of freedom inherent to incommensurate magnetism on the triangular lattice. 

Polycrystalline samples of NiGa$_2$S$_4$ were synthesized by solid state reaction described elsewhere~\cite{nigasStr}.
Elastic-channel data were taken on the GPTAS and HER triple-axis spectrometers at the JRR-3 research reactor.
The backscattering data were taken on the high flux backscattering (HFBS) instrument at the NIST Center for Neutron Research (NCNR)~\cite{nigasNeutron}.
Neutron spin-echo experiments were performed on the NG5-NSE at NCNR and iNSE spectrometers at the JRR-3.
AC susceptibility and nonlinear susceptibilities were measured using a commercial SQUID magnetometer.
Detailed descriptions of experiments, analysis of $\mu$SR data, and discussion can be found in the Supplementary Information.

NiGa$_2$S$_4$ consists of neutral insulating layers held together by the van der Waals force~(Fig.~1(a)).
\begin{figure}[t!]
\includegraphics[width=235pt]{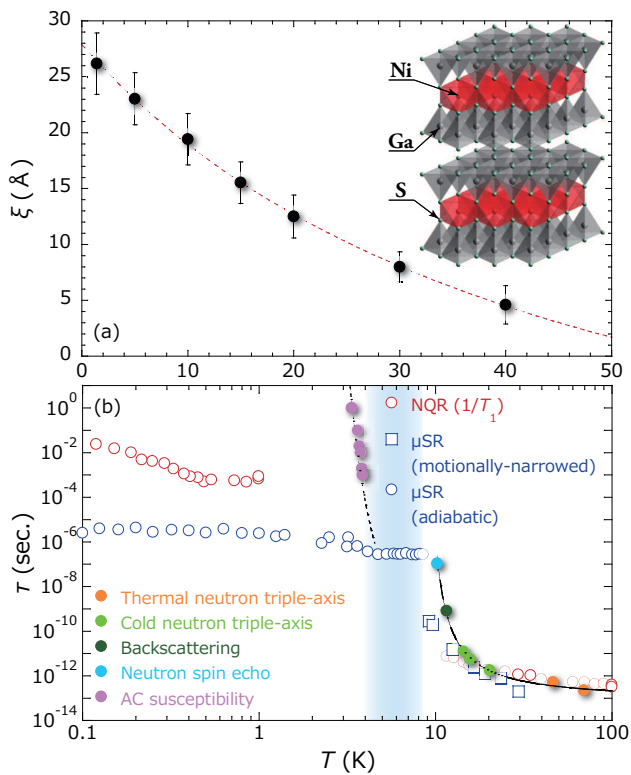}
\caption{(color online). Temperature dependence of (a) the spin correlation length~$\xi$ and (b) characteristic spin relaxation time~$\tau$ of NiGa$_2$S$_4$. The crystal structure of NiGa$_2$S$_4$ is depicted in the inset to (a). NQR data in (b) is after Ref.~\cite{nigasNQR}. Above and below $T^{\ast}$ the muon spin relaxation rates are in the motionally-narrowed ($\propto \tau$) and adiabatic ($\propto 1/\tau$) limits, respectively~\cite{nigasMSR,Zhao12} (see Supplementary Information for details). Vogel-Fulcher behavior from the AC susceptibility measurements is given by the dotted curve. The dashed curve in (a) and the solid curve towards $T^{\ast}$ in (b) are guides to the eyes.}
\end{figure}
This curtails electron hopping between layers and leads to a strongly two-dimensional magnet where spin correlations do not extend beyond nearest-neighbor planes~\cite{nigasNeutron}.
Indeed, a study of the effects of nonmagnetic impurities indicates quantum collective phenomena in NiGa$_2$S$_4$ are controlled by intra-plane interactions~\cite{nigasZndope}.
Within a layer, nickel ions with $S=1$ reside on an equilateral triangular lattice.
NiGa$_2$S$_4$ is thus a rare realization of a highly two-dimensional quantum antiferromagnet.
Upholding expectations of anomalous magnetism, the material does not exhibit conventional magnetic-order down to temperature~$T=50$~mK.
Neutron scattering reveals incommensurate spin correlations with an in-plane correlation length~$\xi$ that gradually increases with cooling but tends to 2.6~nm~(7~lattice spacings) without a thermal anomaly~[Fig.~1(a)].

In sharp contrast, {\it temporal} correlations feature clear thermal anomalies.
Figure~1(b) shows the temperature dependence of the characteristic spin relaxation time~$\tau$ in NiGa$_2$S$_4$ and is the central result of this work.
Unlike normal magnets, NiGa$_2$S$_4$ possesses an intermediate temperature regime with spin dynamics on time scales six orders of magnitude greater than the time scale set by the exchange constant~$J$.
Multiple experimental techniques underlie these data as described in the following.
Gallium nuclear quadrupole resonance~(NQR) experiments~\cite{nigasNQR} and $\mu$SR measurements~\cite{nigasMSR,Zhao12} reveal critical slowing down of Ni-spin fluctuations between the Weiss temperature~$|\theta_{\rm W}|=80$~K and a characteristic temperature $T^{\ast}=8.5$~K, where nominally elastic diffuse magnetic neutron scattering develops.
This is accompanied by a loss of the NQR signal between $T^\ast$ and 2.5(5)~K indicating dynamic nuclear spin relaxation resulting from slow Ni spin fluctuations~\cite{nigasMSR,Zhao12}.
$\mu$SR provides information for $T<T^{\ast}$ where the NQR signal is absent, and shows the spin relaxation rate decreases slightly below $T^{\ast}$ but remains finite down to 25~mK~\cite{Zhao12}.
These results demonstrate Ni spin dynamics persists to the lowest temperatures, with considerable spectral weight in the MHz frequency range.

Figure~2 shows the temperature dependence of nominally elastic magnetic neutron scattering from polycrystalline NiGa$_2$S$_4$ as measured with varying spectrometer energy resolutions.
\begin{figure}[t!]
\includegraphics[width=\linewidth]{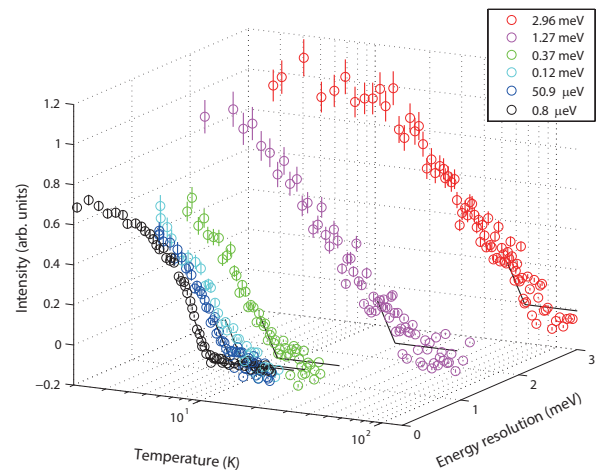}
\caption{(color online). Temperature dependence of the elastic channel data from NiGa$_2$S$_4$. The nominally elastic signals of the magnetic diffuse scattering from polycrystalline NiGa$_2$S$_4$ are obtained using thermal-neutron (energy-resolution $\delta E=$ 2.96, 1.27 meV) and cold-neutron ($\delta E=$ 0.37, 0.12 meV, 50.9 $\mu$eV) triple-axis spectrometers at the wave vector $Q=$ 0.58~{\AA}$^{-1}$, and the backscattering spectrometer ($\delta E=$ 0.8 $\mu$eV) at $Q=$ 0.76~{\AA}$^{-1}$~\cite{nigasNeutron}.}
\end{figure}
The data were obtained on thermal- and cold-neutron triple-axis spectrometers, and backscattering spectrometer~\cite{nigasNeutron}.
The energy-resolution~$\delta E$ sets the minimum spin relaxation time~$\tau\sim\hbar/\delta E$ for spin fluctuations that contribute to intensity in the elastic channel.
Thus the gradual appearance of elastic scattering below a $\delta E$-dependent crossover temperature signals the development of spin correlations on a time scale beyond $\tau\sim\hbar/\delta E$~(Fig.~2)~\cite{Murani}.
The downward shift in onset temperature with tighter energy-resolution indicates the time scale for spin fluctuations increases with cooling as presented in Fig.~1(b).

The smallest value of $\delta E$ is obtained using the neutron spin echo (NSE) technique.
The resulting dependence of the intermediate scattering function~(ISF)~$S(Q,t)/S(Q,0)$ on Fourier time~$t$, is shown in Fig.~3 for three representative temperatures.
\begin{figure}[t!]
\includegraphics[width=210pt]{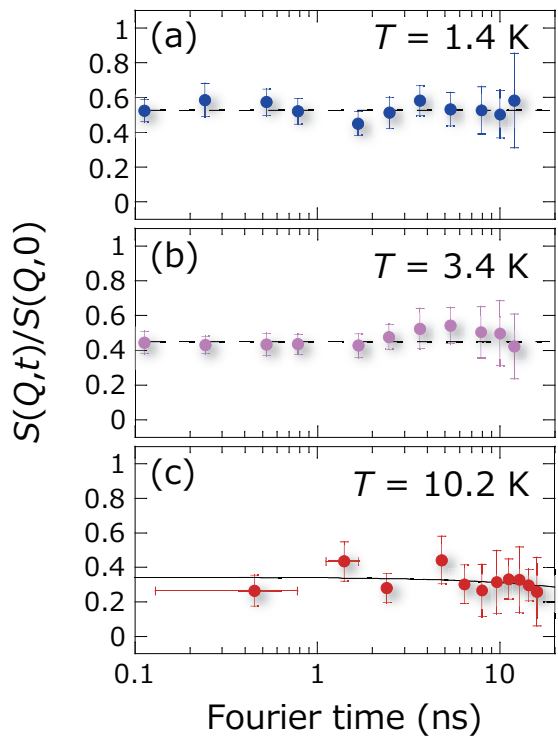}
\caption{(color online). Fourier time~$t$ dependence of the intermediate scattering function~$S(Q,t)/S(Q,0)$ of NiGa$_2$S$_4$. (a) $T=$ 1.4 K data and (b) $T=$ 3.4 K data were acquired on the NG5-NSE instrument at NCNR. (c) $T=$ 10.2 K data are obtained on the C2-3-1-iNSE instrument at JRR-3. All the data were taken at wave vector transfer $Q=$ 0.65~{\AA}$^{-1}$. The dashed lines in (a) and (b) give the averages, and an exponential decay fit is shown as the solid line in (c).}
\end{figure}
To estimate the relaxation time~$\tau$, the data were fit to a simple exponential decay form,~$\exp\left(-t/\tau\right)$.
For data taken at $T=10.2$~K~(Fig.~3(c)) where a slight reduction in $S(Q,t)/S(Q,0)$ is visible up to $t=16$~ns, the fit yields $\tau\sim$~0.11(5)~$\mu$s with reduced $\chi^2=3.971(6)$.
A fit to constant ISF leads to $\chi^2=4.014(6)$.
Figures~3(a) and 3(b) show that below $T^{\ast}$, $S(Q,t)/S(Q,0)$ is constant and does not show any apparent relaxation.
Assuming exponential decay, the data place a lower limit of $\tau=1.3$~$\mu$s on the corresponding relaxation time.
The fact that $S(Q,t)/S(Q,0)<1$ for early times indicates fast initial relaxation and the decrease of $S(Q,t>0.1~{\rm ns})/S(Q,0)$ with increasing temperature~(Fig.~3) indicates a corresponding reduction in the fraction of spins participating in slow relaxation.
Given the temperature dependence of the ISF for Fourier time ranging from 0.05 to 1.0~ns~(Supplementary Information), we interpret the ISF as the volume fraction of spins correlated in accordance with the critical wave vector $Q_c$ and relaxing at times beyond the Fourier time.
The temperature dependence of the time averaged ISF for $0.05\le t\le 1$~ns is given in Fig.~4(a).
Reflecting the behaviour of $\tau$ for $T_0\le T\le T^{\ast}$, where $T_0\sim$~4(1)~K~(Fig.~1(b)), there is a temperature regime where the time-averaged ISF is approximately constant possibly due to quantum tunneling.
The time-averaged ISF however, increases upon cooling below $T_0$, indicating slow dynamics through a larger fraction of the sample, which eventually results in recovery of the NQR signal for $T\le$~2.5(5)~K~\cite{nigasNQR}.

For access to the MHz time scale we turn to $\mu$SR.
A description of the method employed to infer a relaxation time from $\mu$SR data can be found in Supplementary Information.
Interestingly, the relaxation times from resonance data are considerably shorter than inferred from neutron scattering for $T<1.5T^{\ast}$.
This does not however, represent a discrepancy since the techniques probe different aspects of the spin correlations.
In conjunction, the results indicate critical spin-pair correlations probed by neutron scattering at $Q_c$ persist on a longer time scale than the $Q$-averaged correlations probed by NQR and $\mu$SR.

To investigate ultra-slow spin dynamics we use AC susceptibility measurements, which probe the uniform $Q\equiv 0$ response. 
Figure~4(b) gives the in-phase component~$\chi^{\prime}$ as a function of temperature for frequencies from 1 to 1000~Hz.
\begin{figure}[t!]
\includegraphics[width=\linewidth]{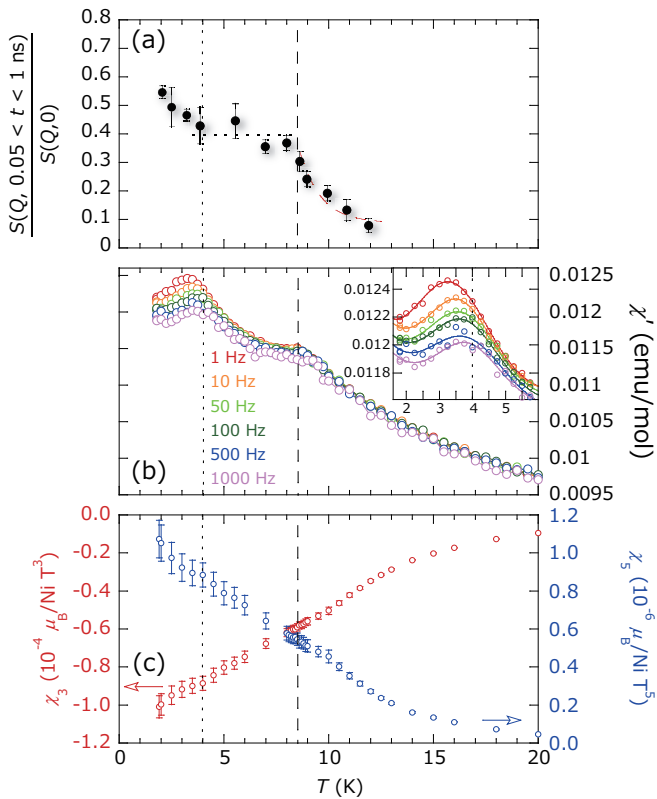}
\caption{(color online). Temperature dependence of (a) the averaged intermediate scattering function~$S(Q,0.05\le t\le 1\ {\rm ns})/S(Q,0)$, (b) AC and (c) nonlinear susceptibilities, the coefficient of the third power (left-axis) and the fifth power (right-axis). The dashed curve and horizontal dotted line in (a) are guides to the eyes. The inset to (b) emphasizes the frequency dependence of $\chi^{\prime}$ at low temperatures. The vertical dashed and dotted lines denote $T^{\ast}=$ 8.5 K and $T_0=$ 4(1) K, respectively.}
\end{figure}
While no significant frequency dependence is observed at the $T^{\ast}$ anomaly where the spin fluctuation rate is in the MHz range, frequency dependence is observed for $T\approx$~3~K where the NQR signal is recovered~\cite{nigasNQR}.
Below 2.5(5)~K, a quasi-static inhomogeneous internal field was inferred from the NQR experiments~\cite{nigasNQR} while $\mu$SR is inhomogeneous at all temperatures~\cite{nigasMSR,Zhao12}.
These observations indicate a wide distribution of spin fluctuation rates, which is corroborated by the distinct time scales inferred from $\mu$SR, NQR and ac susceptibility data (Fig.1(b)).
That this distribution is strongly $Q$-dependent should be expected given the $Q$-dependence of magnetic neutron scattering, and is witnessed by the longer relaxation times for $Q=0$ AC susceptibility and $Q_c$ neutron scattering data compared to local probes~($\mu$SR and NQR).
The relatively slow fluctuations seen in NQR could be responsible for the quasi-static component of the $\mu$SR signal.
Extrapolation of the Vogel-Fulcher curve~(Supplementary Information) for the AC susceptibility shows that microsecond spin dynamics is present at least down to $T_0$.
While neither the $\mu$SR or NSE data exclude a truly static component, the $\mu$SR data indicate significant MHz spin dynamics down to 25~mK~\cite{Zhao12}.

At its freezing temperature, a canonical spin-glass is predicted to have diverging third and fifth-order nonlinear susceptibilities, $\chi_3$ and $\chi_5$~\cite{SG}.
In search of possible spin-glass-type freezing in NiGa$_2$S$_4$, nonlinear susceptibility measurements were performed.
Figures~4(c) shows the temperature dependence of $\chi_3$ and $\chi_5$.
Neither shows divergent behaviour at $T^{\ast}$ or $T_0$.
The quadratic temperature dependence of the low-temperature magnetic specific heat $C_{\rm M}$~\cite{nigasScience} in NiGa$_2$S$_4$ also contrasts with the conventional linear-$T$ behaviour known for spin-glasses.
Similar $C_{\rm M}\sim T^2$ behaviour has been reported for quasi-2D Kagom\'e-related frustrated magnets of deuteronium jarosite~\cite{jarosite} and SCGO~\cite{SCGO}.
The quasi-two-dimensional antiferromagnetic spin-correlations revealed by $Q$-dependent magnetic neutron scattering~\cite{nigasScience,nigasNeutron} differ qualitatively from the canonical spin-glasses, where {\it no} $Q$-dependence is anticipated~\cite{SG}.

Despite the structural simplicity and considerable theoretical efforts, no theory is so far able to fully account for the unusual magnetism of NiGa$_2$S$_4$.
In the following we shall discuss models with potential relevance to the experimental findings.
In Monte-Carlo simulations, classical Heisenberg spins with near-neighbor interactions on the triangular lattice exhibit a precipitous decrease in the spin relaxation rate for $T_{\rm v}/J\sim 0.28$~\cite{SpinGel}, which may be compared to $T^{\ast}/J=0.43$ ($J=20(1)$~K from high-$T$ susceptibility data).
This phenomenon has been associated with topological $Z_2$-vortex ordering~\cite{Z2}.
Slow spin dynamics and a finite correlation length is predicted below $T_{\rm v}$~\cite{SpinGel}.
Aspects of this theory are in accordance with our findings of consecutive anomalies in the magnetic susceptibility and $C_{\rm M}$, and the asymptotic behaviour of $\chi_3$~(Supplementary Information).
However, the correlation length predicted in the low-$T$ regime exceeds that observed in NiGa$_2$S$_4$ by orders of magnitude. 

Lacking in this theory as a model for NiGa$_2$S$_4$ is the quantum nature of Ni-spins and the incommensurate $Q_c$.
The semi-classical description of spin degrees of freedom thus cannot account for the impurity-spin size dependence of doped NiGa$_2$S$_4$~\cite{nigasZndope,nigasImpurity1,nigasImpurity2}.
Specifically, the low-temperature $T^2$ asymptotic behaviour of $C_{\rm M}$ is preserved only for integer impurity spins that support a quadrupole moment~\cite{nematic1,nematic2}.
Indeed, a recent theory achieves slow spin dynamics in a near-neighbor model of NiGa$_2$S$_4$ with impurity spins induced by bond-disorder with long-range interactions mediated by an antiferroquadrupolar bulk phase~\cite{nematic-imp}.

None of these models however, considers the incommensurate nature of spin correlations in NiGa$_2$S$_4$ which are thought to arise from competing ferromagnetic nearest-neighbor and antiferromagnetic third near-neighbor interactions~\cite{nigasScience}.
Apart from global rotations, the conventional 120 degree spin structure on the triangular lattice comes in two flavours distinguished by the scalar chirality of the three spins on a specific triangle.
However, the incommensurate structure of NiGa$_2$S$_4$ has three different wave-vector domains, which are related to each other by a $C_3$ rotation.
Monte-Carlo simulations show that $C_3$ symmetry breaking occurs through a finite temperature first-order phase transition~\cite{C3-kawashima,C3-balents}.
Based on the inferred exchange interactions, the predicted critical temperature~($0.24J_3\sim$~6~K) is not far from $T^{\ast}$.
The Imry-Ma criterion~\cite{ImryMa} however, implies this transition does not survive under arbitrary weak random field disorder.
$T^{\ast}$ may thus be the remnant of this first-order transition and the anomalous MHz dynamics might arise from interfaces between the associated $C_3$-domains~\cite{tomita}.

We have characterized megahertz dynamics in a short-range correlated state finding a temperature dependent relaxation time that is not reflected in temperature dependent spatial correlations.
A spin state with emergent topological degrees of freedom that form at $T^{\ast}$ and freeze at lower-$T$ may support decoupling between {\it spatial} and {\it temporal} two-point spin correlations.
Low frequency~($\hbar\omega\ll J$) and low temperature~($k_BT< J$) spin fluctuations as reported here are not uncommon in geometrically frustrated magnets but they are poorly understood.
Owing to its structural simplicity and the availability of high quality single crystals, NiGa$_2$S$_4$ presents an excellent opportunity for progress towards understanding slow dynamics in spin systems with short-range correlations.

We are grateful for fruitful discussions with P. Dalmas de R\'eotier, K. Ishida, 
Y. Maeno, S. Onoda, O. Tchernyshyov, H. Tsunetsugu, J.-J. Wen and J.~D. Zang.
This work was partially supported by Grants-in-Aid for Scientific Research from Japan Society for the Promotion of Science (JSPS) (Nos.~24740223 and 25707030), and by PRESTO of JST\@, and by Program for Advancing Strategic International Networks to Accelerate the Circulation of Talented Researchers (No. R2604) from the JSPS.
Work at IQM was supported by U.S. Department of Energy, Office of Basic Energy Sciences, Division of Materials Sciences and Engineering under award No.~DE-FG02-08ER46544. 
Work at NIST was supported in part by the National Science Foundation under Agreement No.~DMR-0944772. 
Work at U.C.~Riverside was supported by the U.S. National Science Foundation, Grant No.~0801407.

\end{document}